\documentclass[aps,prb,reprint,groupedaddress]{revtex4-1}
\usepackage{graphicx}
\usepackage{xspace}
\newcommand{\baco}{BaCo$_{\text{2}}$V$_{\text{2}}$O$_{\text{8}}$\xspace}
\newcommand{\srco}{SrCo$_{\text{2}}$V$_{\text{2}}$O$_{\text{8}}$\xspace}
\newcommand{\amo}{$AM_{\text{2}}$V$_{\text{2}}$O$_{\text{8}}$\xspace}

\newcommand{\tn}{$T_{\text{N}}$\xspace}

\begin{document}

\title{Magnetic phase diagrams, domain switching and a quantum phase transition of the quasi-1D Ising-like antiferromagnet BaCo$_{\text{2}}$V$_{\text{2}}$O$_{\text{8}}$}

\author{S.~K.~Niesen$^1$}
\author{G.~Kolland$^1$}
\author{M.~Seher$^1$}
\author{O.~Breunig$^1$}
\author{M.~Valldor$^1$}
\author{M.~Braden$^1$}
\author{B.~Grenier$^2$}
\author{T.~Lorenz$^1$}
\email[]{tl@ph2.uni-koeln.de}
\affiliation{$^1$II.~Physikalisches Institut, Universit\"{a}t zu
K\"{o}ln, Z\"{u}lpicher Stra{\ss}e 77, 50937 K\"{o}ln, Germany \\
$^2$SPSMS, UMR-E 9001, CEA-INAC/UJF-Grenoble I, Laboratoire MDN, 17 rue des martyrs, 38054 Grenoble cedex 9, France}

\date{\today}

\begin{abstract}

In the effective Ising spin-1/2 antiferromagnetic chain system \baco, the magnetic-field influence is highly anisotropic.  
For magnetic fields along the easy axis $c$, the N\'{e}el order is strongly suppressed already for low fields and an incommensurate order is entered above 4~T. We present a detailed study of the magnetic phase diagrams for different magnetic field directions, which are derived from magnetization data, high-resolution thermal expansion and magnetostriction measurements as well as from the thermal conductivity.
Zero-field thermal expansion data reveal that the magnetic transition is accompanied by an orthorhombic distortion within the $ab$ plane. Under ambient conditions the crystals are heavily twinned, but the domain orientation can be influenced either by applying uniaxial pressure or a magnetic field along the [100] direction. In addition, our data reveal a pronounced in-plane magnetic anisotropy for fields applied within the $ab$ plane. For $H \| [110]$, the magnetic field influence on \tn is weak, whereas for magnetic fields applied along [100], \tn vanishes at about 10~T and the zero-field N\'{e}el order is completely suppressed as is confirmed by neutron diffraction data. The second-order phase transition strongly suggests a quantum critical point being present at $H\simeq 10$~T parallel [100], where the N\'{e}el order probably changes to a spin-liquid state.

\end{abstract}


\maketitle

\section{Introduction}
\label{sec:Intro}

Low-dimensional quantum spin systems are of particular interest in solid-state research due to their anomalous ground-state properties and unusual excitations, which are strongly influenced by enhanced quantum fluctuations. \baco belongs to the material class \amo , which contains screw chains of $M$O$_6$ octahedra running along the $c$ axis of the tetragonal crystal structure~\cite{Wichmann1986}; space group \textit{I}4$_1$/\textit{acd} with lattice constants  $a=12.444(1)$~{\AA} and $c=8.415(3)$~{\AA}. In the $ab$ planes, the chains are separated by nonmagnetic $A^{2+}$ ($A=$~Sr, Ba) and V$^{5+}$ ions. Depending on the transition metal ion $M^{2+}$ ($M=$~Cu, Ni, Co, Mn), the various members of \amo represent spin-chain materials with different spin quantum numbers and anisotropies~\cite{muebuwich, muebu92, muebu94, PhysRevB.69.220407, 2007JSSCh.180.1770H, Niesen2011}. According to Hund's first rule, the Co$^{2+}$ ions realize the $S=3/2$ high-spin state and the lower-lying $t_{2g}$ orbitals of the crystal-field split $3d$ states are only partially filled, resulting in a finite effective orbital moment $\tilde{l}=1$ for regular CoO$_6$ octahedra. In \baco, the CoO$_6$ are compressed along the $c$ axis, which causes a pronounced Ising anisotropy of the total magnetic moment with $c$ as the easy axis. This Ising anisotropy is experimentally confirmed by low-temperature high-field magnetization data~\cite{Kimura2006} which reveal  a strong anisotropy of both, the saturation field $H_{sat}^{\|c} \approx 0.5 \, H_{sat}^{\perp c}$ and the saturation magnetization $M_{sat}^{\|c} \approx 1.5\, M_{sat}^{\perp c}$, as well as by temperature-dependent magnetic-susceptibility measurements~\cite{Hee2005} in the intermediate temperature range yielding $\chi^{\|c}(T) \approx 2 \chi^{\perp c}(T)$ for $T> 5.5$~K. Moreover, $\chi^{\|c}(T)$ shows a pronounced maximum around 35~K, which signals short-range magnetic correlations reflecting the one-dimensional nature of the magnetic subsystem due to the dominant magnetic exchange along the CoO$_6$ screw chains. Thus, \baco represents a model system of a (quasi-)one-dimensional antiferromagnetic effective Ising $S=1/2$ chain that is described by the XXZ hamiltonian 
\begin{equation}
\mathcal{H}= \sum_i J\left\{S_i^z S_{i+1}^z + \varepsilon \left(S_i^x S_{i+1}^x+S_i^y S_{i+1}^y  \right) \right\} +   g \mu_B {\bf S}_i {\bf H} \, .
\label{hamil}
\end{equation}
From an analysis of the low-temperature magnetization data, $J\simeq 65$~K and $\varepsilon \simeq 0.46$ as well as the anisotropic $g$ factors $g^{\|c}\simeq 6.3$ and $g^{\perp c}\simeq 3.2$ have been derived~\cite{Kimura2006}.

Due to finite inter-chain couplings, the system shows long-range antiferromagnetic order below $T_{\text{N}}\approx  5.5$~K with a  propagation vector $\vec{k}_{\text{AF}}=(1,0,0)$ and the spin direction along the $\pm c$ axis. Nearest-neighbor (NN) spins along $c$ are oriented antiparallel to each other, while within the (001) planes, NN spins point either in the same or in the opposite direction \cite{Kimura2008A, Kawasaki2011,Grenier2012}. As a consequence, the magnetic symmetry is less than tetragonal and the ordered phase typically contains different domains, which are rotated by 90$^\circ$ and translated by $c/4$ with respect to each other~\cite{Grenier2012}. The magnetic-field $vs.$ temperature phase diagram of \baco has been studied by several techniques~\cite{He2006A,Kimura2008,Kimura2008A,Kimura2009,Kimura2010,Yamaguchi2011,Kawasaki2011,Klanjsek2012, Grenier2012, Zhao2012} and a pronounced anisotropy has been reported~\cite{He2006A, Zhao2012}. For $H\| c$, \tn rapidly decreases in the field range up to about 4~T and above this field an incommensurate (IC) magnetic phase with a weakly field-dependent transition temperature $T_{\text{IC}}\approx  1$~K develops. Moreover, experimental evidence for additional phase transitions in the region of the IC phase has been reported~\cite{Yamaguchi2011, Klanjsek2012}. For $H\perp c$, a rather weak decrease  $\partial T_{\text{N}}/\partial H \approx -0.1$~K/T has been reported up to a maximum field of 9~T in Ref.~\onlinecite{He2006A}, whereas a much stronger decrease of  $\partial T_{\text{N}}/\partial H \approx -1 $~K/T for $8 < H < 10$~T has been found in Ref.~\onlinecite{Yamaguchi2011}. However, in most of the published studies for $H\perp c$, the explicit direction of the magnetic field within the $ab$ plane has not been  given~\cite{Hee2005,He2005,Kimura2006,He2006A,Kuo2009,Lejay2011,Yamaguchi2011}. 

In this report, we present a detailed study of the magnetic phase diagrams of \baco for different directions of the magnetic field. Based on high-resolution thermal-expansion measurements by capacitance dilatometry, we find that the magnetic ordering in zero magnetic field is accompanied by small structural distortions, which break the tetragonal symmetry. The distortions arise from the magnetoelastic couplings within the $ab$ plane and the crystals are twinned according to the magnetic domain structure. For $H \|c$, our data confirm the existing phase diagram in the low-field range and reveal a nonmonotonic field dependence of $T_{\text{IC}}$ towards higher fields. For $H \perp c$, we find an additional anisotropy, which has been overlooked so far and resolves the apparent discrepancy between the phase diagrams of Refs.~\onlinecite{He2006A, Zhao2012}. For a magnetic field applied along the [110] direction, we observe, in fact, the above-mentioned weak decrease of $\partial T_{\text{N}}/\partial H \approx -0.1$~K/T up to our maximum field of 16~T, but for $H \| [100]$ (respectively $\| [010]$ of the twinned crystals), a much stronger decrease of \tn\ occurs (in agreement with the data of Ref.~\onlinecite{Zhao2012}) and our data suggest a quantum phase transition at $H_c\simeq 10$~T. This additional in-plane anisotropy of the magnetic-field influence has been also reported by a very recent magnetization study~\cite{Kimura2013}. Moreover, our measurements reveal that the crystal can be (at least partially) detwinned by applying either uniaxial pressure or a magnetic field along [100]. The paper is organized as follows: Details of the single crystal growth and the experimental setups are described in  section~\ref{sec:exp}, while the experimental results are presented and discussed in section~\ref{sec:ResDis}, which is split into 4 subsections discussing the zero-field measurements and the results for the different field directions $H \|c$, $H \| [110]$, and $H \|a$, respectively.

\section{Experimental}
\label{sec:exp}
\begin{figure}
	\centering
		\includegraphics[width=0.90\linewidth]{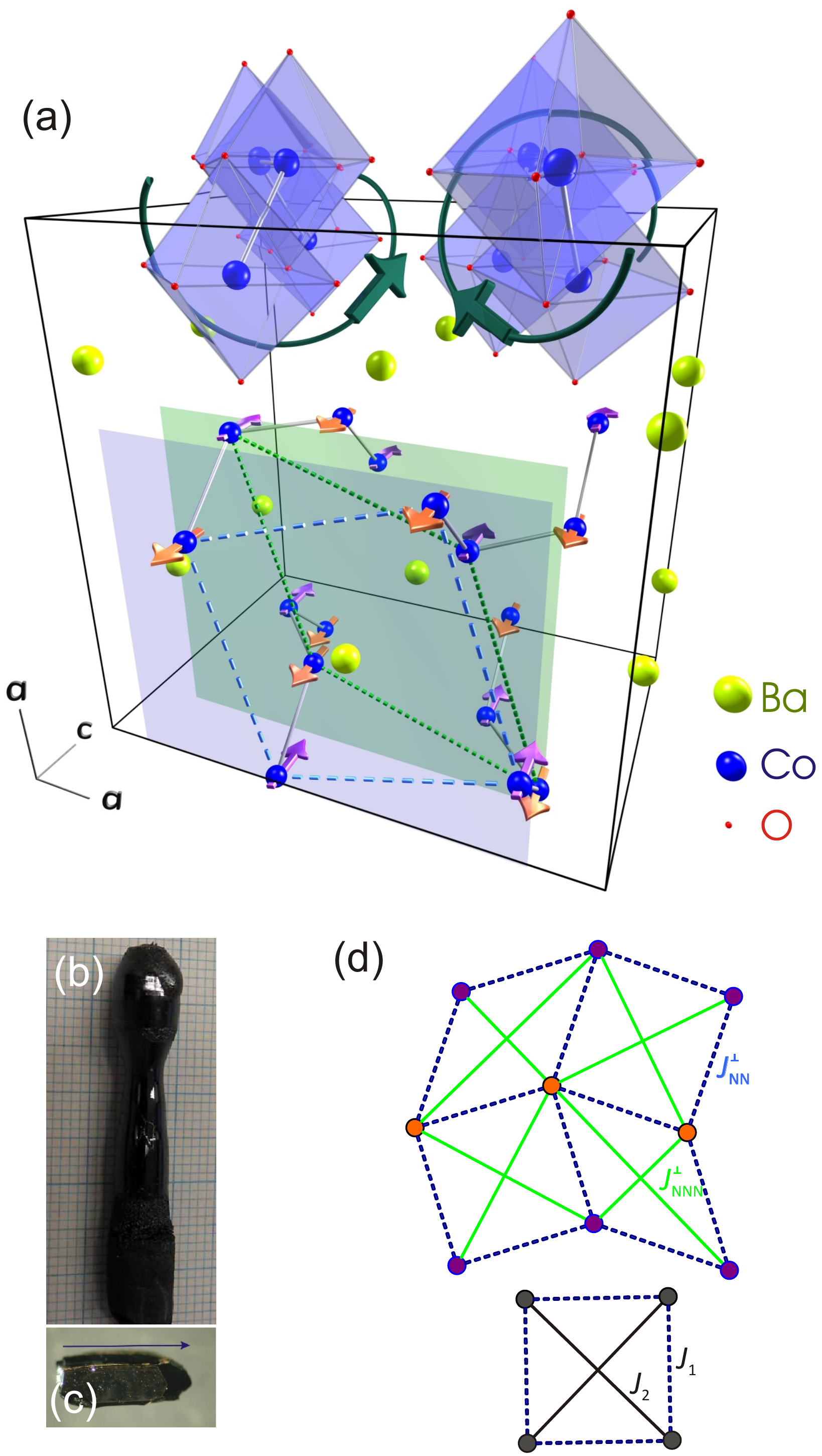}
	\caption{Single crystals of BaCo$_2$V$_2$O$_8$ obtained by the floating-zone method (b) and  by spontaneous nucleation (c). The arrow in (c) marks the magnetic easy $c$ axis and the flat surface is a (110) plane. Panel~(a) summarizes basic features of the crystal and magnetic structure. Right- and left-handed screw chains of CoO$_6$ octahedra are running along $c$. For clarity, four of these chains are also shown without the O$^{2-}$ ions and the dominating NN exchange $J$ along the chains, see Eq.~(\ref{hamil}), is depicted by solid lines. The dashed and dotted lines connect NN Co$^{2+}$ ions in the $ab$ planes	 at $z=1/8$ and $z=3/8$, respectively, and represent NN inter-chain couplings $J^\perp_{NN}$, which compete with the corresponding NNN inter-chain couplings $J^\perp_{NNN}$. Panel (d) displays the geometrical arrangement of the Co$^{2+}$ ions in the $ab$ planes. Due to the competing inter-chain couplings $J^\perp_{NN}$ and $J^\perp_{NNN}$, this arrangement represents a more complex version of the $J_1-J_2$ model on a square lattice~\cite{remark,chandraPhysRevB.38.9335,MelziPhysRevLett.85.1318,Klanjsek2012}.}
	\label{fig:Foto}
\end{figure}

Single crystals of \baco have been synthesized by spontaneous nucleation as well as by the floating-zone method.
In both cases, \baco was synthesized by a solid-state reaction using a mixture of BaCO$_3$ (99+~\% Merck), Co$_3$O$_4$ (99.5~\% Alfa Aesar) and V$_2$O$_5$ (99.5~\% Strem Chemicals) in a 1:$\frac{2}{3}$:1 molar ratio. 
A four-mirror image furnace (FZ-T-10000-H-VI-VP, Crystal Systems Inc.) was used for the single-crystal growth by means of the floating-zone technique (in analogy to Ref.~\onlinecite{Lejay2011}). Atmospheric air at ambient pressure served as reaction atmosphere. Fig.~\ref{fig:Foto} shows photographs of both crystals. The single crystal grown by spontaneous nucleation is of rectangular shape with dimensions of about $1.2 \times 1.3 \times 2.1$~mm$^3$. The long axis is parallel to the $c$ direction and the flat surfaces are (110) planes.

The magnetization has been measured in a vibrating sample magnetometer (PPMS, Quantum Design) for temperatures above 2~K up to a maximum magnetic field of 14~T. Temperature or magnetic-field induced relative length changes $\Delta L(T,H)/L$, \textit{i.e.}\ thermal expansion and magnetostriction, have been determined by a home-built capacitance dilatometer in the temperature range from about 250~mK to 10~K in magnetic fields up to 16~T. Moreover, zero-field thermal-expansion measurements have been performed in another dilatometer, which allows to easily vary the weak uniaxial pressure applied by the springs connecting the moveable and fixed capacitor plates. The corresponding expansion coefficients $\alpha(T)=1/L\, \partial \Delta L(T)/\partial T$ and $\lambda(H)=1/L\, \partial \Delta L(H)/\partial H$ have been obtained numerically. In finite fields, the length change has been determined parallel to the direction of the applied field, \textit{i.e.}\ $L_i \| H_i$.  The thermal conductivity has been measured by the standard steady-state method from $\simeq 0.3$ to 15~K. The temperature gradient was produced by a heater at one end of the sample and measured by two matched RuO$_2$ thermometers. The directions of the heat current were either along the $c$ axis or the [110] direction of the crystal of Fig.~\ref{fig:Foto}c and magnetic fields up to $\simeq 15$~T have been applied along $c$, along $a$, and along the [110] direction. Neutron diffraction studies were performed at $T\simeq 50$~mK in magnetic field up to 12~T on the CEA-CRG single-crystal D23 diffractometer, operated in the normal beam mode with an incident wavelength of 1.2765 Å. The sample of about 250~mm$^3$ was cut from the crystal rod obtained by the floating-zone method and was aligned with the $a$ axis vertical. It was mounted in a 12~T vertical cryomagnet equipped with a dilution insert, with the field applied along the $a$ axis.

\section{Results and Discussion}
\label{sec:ResDis}

\subsection{Zero-field thermal expansion}

\begin{figure}[t]
	\centering
		\includegraphics[width=1.00\linewidth]{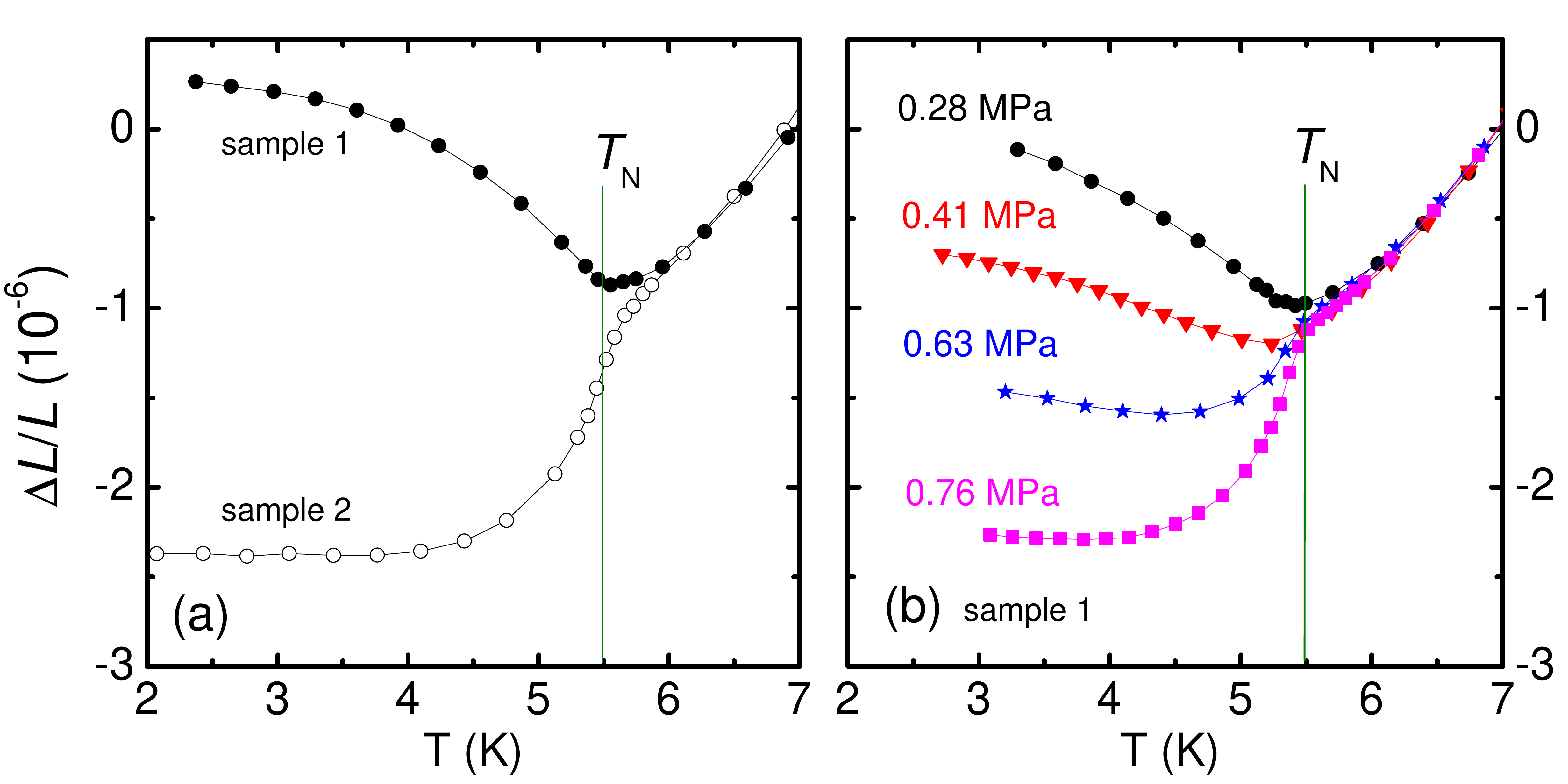}
		\includegraphics[width=.9\linewidth]{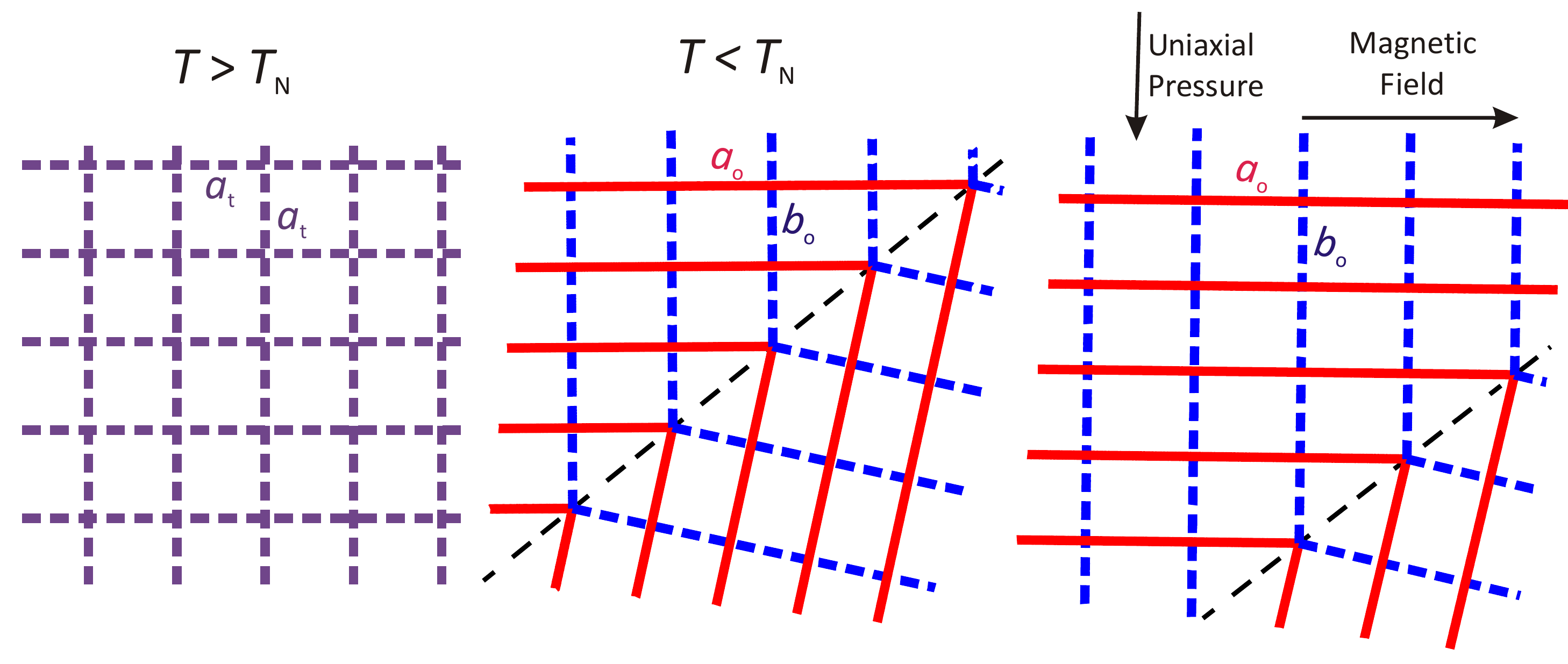}
		\caption{(color online) Zero-field thermal expansion of the tetragonal $a_t$ axis of \baco. The different data sets in (a) were obtained on two different setups and samples, while those in (b) were measured during the same run, but with varying uniaxial pressure $p_i$ applied along $a_t$. All length changes $\Delta L=L(T)-L_0$ are related to $L_0=L(7~{\rm K}, p_i)$. The process of a partial detwinning by uniaxial pressure or in a magnetic field is schematically sketched in the lower panel. }
	\label{fig:nullfeld}
\end{figure}

Fig.~\ref{fig:nullfeld} displays thermal-expansion data along the $a$ axis of \baco measured in zero magnetic field. The left panel shows  $\Delta L(T)/L$ obtained on two different samples in different setups. In both cases, the magnetic ordering is accompanied by spontaneous strains, but these strains are of different signs. Because tetragonal symmetry requires an isotropic thermal expansion within the $ab$ plane, this observation is an unambiguous proof that the tetragonal symmetry is broken in the N\'{e}el-ordered phase. This can be naturally explained by assuming that the antiferromagnetic ordering causes a small orthorhombic splitting $a\neq b$ below \tn and that the crystals are twinned with respect to the orientations of $a$ and $b$. Consequently, the measurements of Fig.~\ref{fig:nullfeld}a reflect different superpositions of the $a$- and $b$-axis expansions due to different twin ratios in both runs. In order to check this, we performed a series of thermal-expansion measurements along the tetragonal $a_t$ axis of one crystal for varying uniaxial pressure along this axis. The uniaxial pressure $p_i$ arises from the spring which fixes the sample and the moveable plate to the fixed capacitor plate and increases with increasing base capacitance $C_0\propto 1/d_0$, where $d_0$ is the base distance between both capacitor plates. As the spring of the actual setup is rather soft, the resulting $p_i$ for a sample with a cross section in the mm$^2$ range can be varied around $\approx 0.5$~MPa. 

As is shown in Fig.~\ref{fig:nullfeld}b, increasing $p_i$ from 0.28~MPa to 0.76~MPa is indeed sufficient to switch from a spontaneous expansion below \tn to a spontaneous contraction.  In this low-pressure range, the magnitude of the spontaneous strain continuously varies with increasing $p_i$, what is naturally interpreted  as follows. Below $T_{\rm N}$, the tetragonal $a_t$ axes spontaneously develop an orthorhombic splitting, \textit{e.g.}\ $a_o > a_t > b_o$, but due to a  twinning of the ferroelastic domains, the measured macroscopic length change is an average of $a_o(T)$ and $b_o(T)$. With increasing $p_i$ along one of the tetragonal $a_t$ axes, the amount of $b_o$-oriented  domains along this direction is continuously enhanced until a single-domain crystal would be reached. This process is schematically sketched in Fig.~\ref{fig:nullfeld}c, but the uniaxial pressure applied here is not sufficient to reach this single-domain limit. The detwinning process should be directly visible as a function of $p_i$ below \tn, but similar to the low-temperature magnetization [polarization] loops $M(H)$[$P(E)$] in ferromagnetic [ferroelectric] materials, pronounced hysteresis effects may occur. Therefore (and because of technical reasons in our setup), we performed ``pressure-cooled'' measurements meaning that the uniaxial pressure has been changed well above \tn at $\simeq 10$~K, then the sample has been cooled to $\simeq 3$~K and the data of Fig.~\ref{fig:nullfeld}b have been collected during the subsequent heating run. As will be discussed below (Section~\ref{sec:Hparallela}), an analogous detwinning is also achieved by applying a magnetic field along $a_t$ and in this case, we can also observe the typical hysteresis effects of zero-field cooled measurements. 

The structural symmetry reduction in the N\'{e}el-ordered phase of \baco has not been discussed in any of the reported diffraction studies~\cite{Hee2005,Kimura2008A,Kimura2010,Kawasaki2011,Grenier2012}, what is most probably due to the fact that the orthorhombic splitting is very small. As shown in Fig.~\ref{fig:nullfeld}, the averaged macroscopic strains  $\Delta L/L$ are of the order of $10^{-6}$ and as will be discussed below, we expect that even in a completely detwinned sample $\Delta L/L$ does not significantly exceed $10^{-4}$. Thus, it will be hard to detect such small changes by diffraction techniques. Nevertheless, our finding is supported by the magnetic structure obtained by neutron scattering~\cite{Kawasaki2011,Grenier2012}. As is shown in Fig.~\ref{fig:Foto}, NN spins in the $ab$ plane are oriented parallel to each other along one tetragonal $a_t$ axis, but antiparallel to each other along the other $a_t$ axis. This magnetic structure typically arises from magnetic frustration, \textit{e.g.}\ a competition between antiferromagnetic NN interactions $J^\perp_{NN}$ along $a_t$ and antiferromagnetic interactions $J^\perp_{NNN}$ between the next-nearest neighbors  along the diagonal $\langle 110 \rangle$ directions~\cite{remark}; see Fig.~\ref{fig:Foto} and the corresponding discussions in  Refs.~\onlinecite{chandraPhysRevB.38.9335,MelziPhysRevLett.85.1318,Klanjsek2012,ValldorBaMn2O3,KomarekVOCl,XuJ1J2FeAs}. Here, the notations $J^\perp_{n}$ are used because we are discussing inter-chain interactions, whereas the dominant magnetic intra-chain interaction $J$ of the hamiltonian~(\ref{hamil}) acts along the chain direction.  The competition between  $J^\perp_{NN}$ and $J^\perp_{NNN}$ results in a perfect magnetic frustration for a tetragonal symmetry where both $a_t$ directions are equivalent, but it can easily be lifted by magnetostriction if there is a finite magnetoelastic coupling, \textit{i.e.} if  $J^\perp_{NN}$ and/or  $J^\perp_{NNN}$ change with lattice distortions. One of the simplest scenarios would be that $J_{NN}$ decreases with increasing NN distance.  In this case, a small orthorhombic distortion $a_o > a_t > b_o$ lifts the magnetic frustration and couples the magnetic and lattice structures such that NN spins are parallel to each other along $a_o$ and antiparallel along $b_o$. Note that this oversimplified model could already qualitatively explain the above observation, but at present, neither the sign nor the magnitude of the change of $J_{NN}$ as a function of the distance is known and, moreover,  $J_{NNN}$ can also depend on the lattice distortions.

\subsection{Magnetic field parallel to the \textit{c} axis}

\begin{figure}[t]
	\centering
		\includegraphics[width=1.\linewidth]{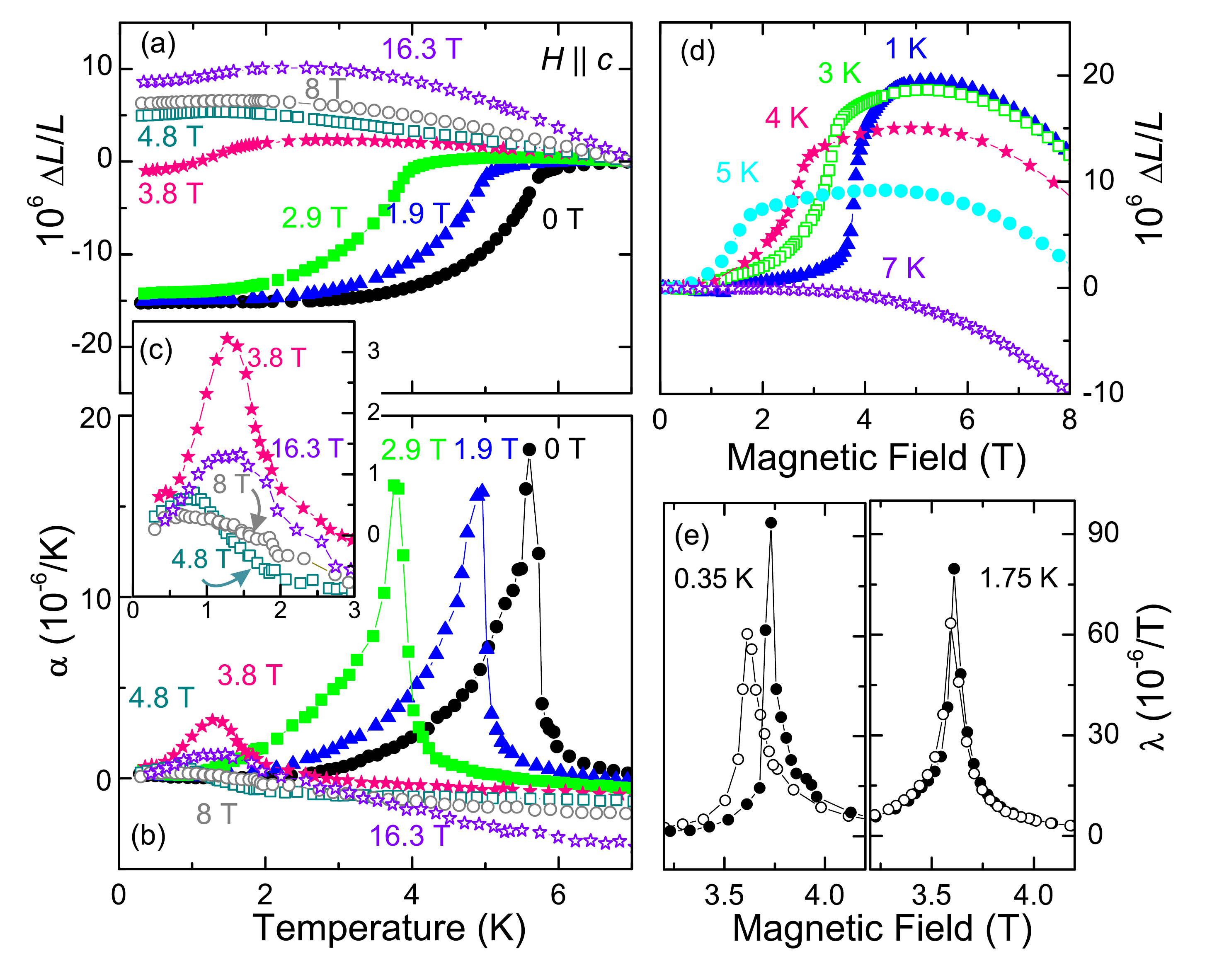}
	\caption{(color online) Thermal expansion (left) and magnetostriction (right) of the $c$ axis of \baco for magnetic fields $H\|c$. In the  upper panels the relative length changes (related to $L_0=L(7~{\rm K}, H)$ and $L_0=L(T, 0~{\rm T})$, respectively) are shown, whereas the lower panels display the derivatives of $\Delta L/L$ with respect to $T$ or $H$. Panel (c) is an enlarged view of the low-temperature range of (b); the different curves in (e) are measured with increasing ($\bullet$) or decreasing ($\circ$) field. 
}
	\label{fig:tadc}
\end{figure}

The thermal-expansion and magnetostriction data along the chain direction $c$ for different magnetic fields $H\| c$, \textit{i.e.}\ along the Ising axis, are summarized in Fig.~\ref{fig:tadc}. The N\'{e}el ordering in low fields is accompanied by a pronounced spontaneous contraction (Fig.~\ref{fig:tadc}a) and the field-dependent $T_{\rm N}(H)$ can be easily monitored by considering the shift of the corresponding $\alpha(T)$ anomalies (Fig.~\ref{fig:tadc}b). The shape and magnitude of these anomalies hardly change up to about 3~T, but with further increasing field the anomalies rapidly drop in size and drastically broaden. As shown in Fig.~\ref{fig:tadc}c, the position as well as the magnitude of these anomalies non-monotonically vary with increasing $H$. The drastic change of the anomalies around 4~T is related to a transition from the N\'{e}el-ordered state to an incommensurate high-field phase as has been established by diffraction studies~\cite{Kimura2008A,Grenier2012}. The corresponding length change have also been studied by low-temperature magnetostriction measurements $\Delta L(H)/L$, see Fig.~\ref{fig:tadc}d. Below about 1.5~K, the length changes at $\approx 4$~T are almost discontinuous, and depending on the field-sweep direction, there is a clear hysteresis confirming the proposed $1^{st}$-order character of this transition (Fig.~\ref{fig:tadc}e). Above 1.5~K, the magnetostriction curves are more continuous and the hysteresis vanishes, \textit{i.e.} the transition is of $2^{nd}$ order. 

At least qualitatively, all the features of the field- and temperature-dependent length changes $\Delta L(H,T)$ of \baco\ strongly resemble the spontaneous strains observed in the spin-Peierls compound CuGeO$_3$~\cite{lorenz97,lorenz98,buechner99}. In both materials, the spontaneous strains are large in the respective low-field commensurate phase and become considerably reduced when an incommensurate high-field phase is entered. Moreover, the temperature-induced transitions into the incommensurate phases are broadened compared to the transitions into the commensurate phases. For doped CuGeO$_3$, it has been observed that the degree of broadening is enhanced with increasing doping concentration suggesting that the incommensurate phase is more severely disturbed by disorder than the commensurate phase~\cite{buechner99}. Because the field-induced broadening in pure \baco is already more pronounced than that in weakly doped CuGeO$_3$, one may speculate that the rather complex structure of \baco could be very sensitive to defects and this could be the reason for the drastic broadening of the transitions to the incommensurate phase. However, the spin-Peierls order in CuGeO$_3$ can obviously not be directly compared to the N\'{e}el order of \baco where the microscopic nature of the high-field phase is not yet clear (see below). Nevertheless, the mere fact that there is a pronounced spontaneous contraction of $c$ in the N\'{e}el-ordered phase establishes a significant magnetoelastic coupling. Again, we consider the most simple scenario of a NN interaction $J$, see Eq.~(\ref{hamil}), which increases with decreasing NN distance. In this case, the spontaneous contraction $\Delta c$ below \tn would result from a gain in magnetic energy ($E_{mag} \propto -\Delta c$), which overcompensates the corresponding loss of elastic energy ($E_{el} \propto \Delta c^2$). Together with the above discussion of the orthorhombic distortions $a_o$ and $b_o$, we can, thus, summarize that the perhaps most simple model of antiferromagnetic NN intra-chain ($J$) and inter-chain ($J^\perp_{NN}$) interactions can explain the observed lattice distortions below \tn . We emphasize, however, that this scenario is oversimplified because (i) the considered NN couplings $J^i\|i$ do not only depend on the corresponding longitudinal distortions $\|i$, but also on the distortions in the transverse directions $j,k\perp i$ and (ii) additional, \textit{e.g.} NNN, couplings and their dependencies on distortions along $i,j,k$ may also be important for a quantitative description.       

\begin{figure}[t]
	\centering
		\includegraphics[width=0.98\linewidth]{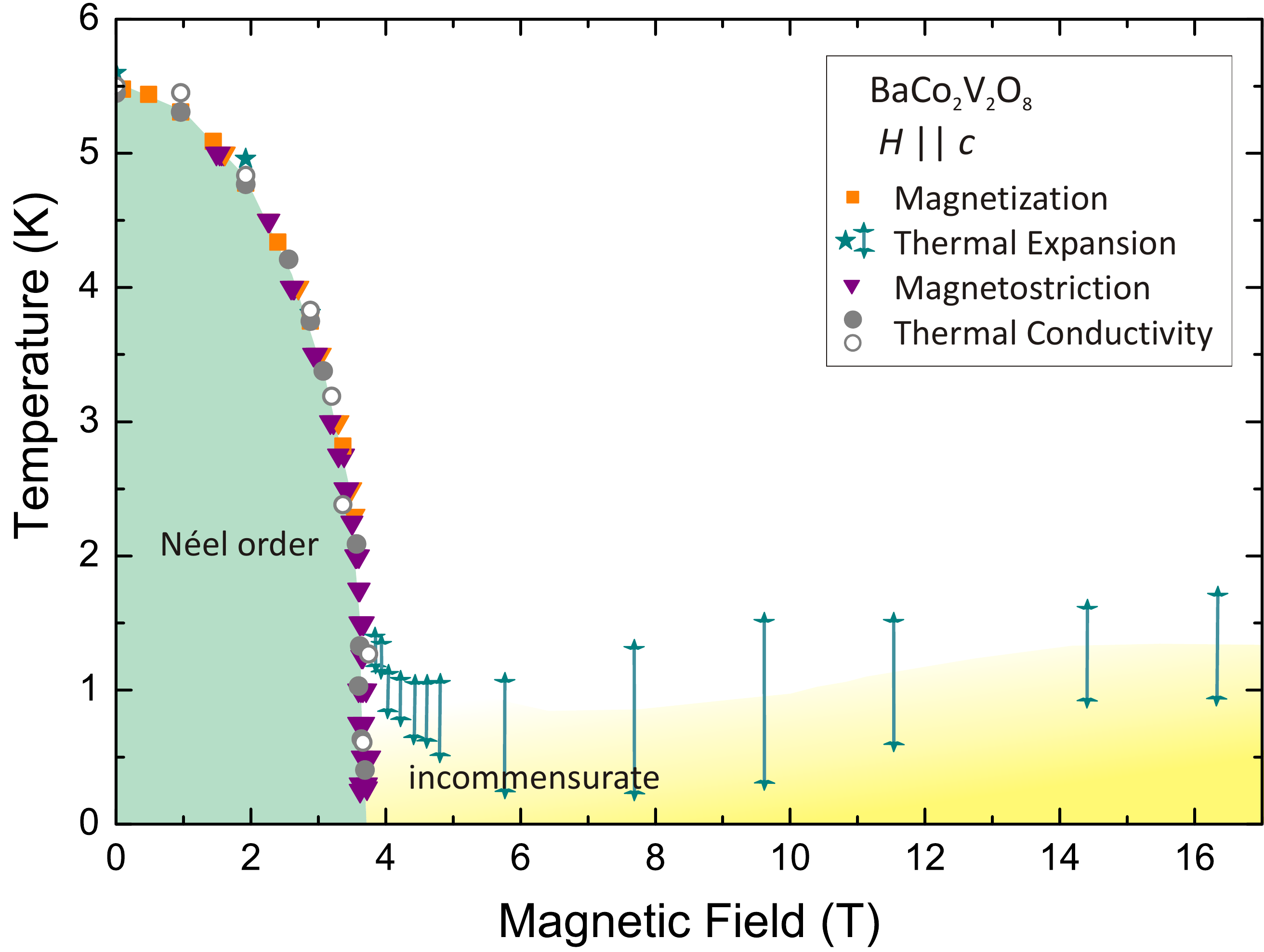}
	\caption{(color online) Phase diagram of \baco for $H\| c$. While the transitions to the N\'{e}el-ordered phase are very sharp, the vertical bars in the higher-field range mark broad crossover regions to the incommensurate phase.}
	\label{fig:phadic}
\end{figure}

In Fig.~\ref{fig:phadic}, we present  the corresponding phase diagram of \baco\ for magnetic fields along the $c$~axis. The phase boundaries are obtained from the positions of the extrema of $\alpha(T)$ and $\lambda(H)$.  Additional data points based on our magnetization and thermal-conductivity measurements are also included. In all cases, the anomalies at the transitions to the N\'{e}el-ordered phase are rather sharp, while those related to the incommensurate phase are very broad. Because the high-field anomalies rather indicate a broad crossover behavior instead of well-defined phase transitions, we marked the positions of the maxima of $\alpha(T)$ in the phase diagram of Fig.~\ref{fig:phadic} as vertical bars representing the anomaly width. The phase diagram agrees well to those presented previously by using other techniques~\cite{He2005,Kimura2008A, Kimura2008,Kimura2009,Yamaguchi2011,Klanjsek2012,Grenier2012}. In the low-field range of the N\'{e}el-ordered phase all these reports essentially agree with each other, but some discrepancies are present in the high-field range. For example,  additional features have been observed in sound-velocity measurements in the field range close to the triple point~\cite{Yamaguchi2011}, but corresponding effects are neither seen in the other literature data nor in our measurements. Most of the previous reports could follow the boundary to the high-field phase only up to $\simeq 10$~T~\cite{Kimura2008,Kimura2008A,Kimura2009,Kimura2010,Yamaguchi2011,Kawasaki2011,Grenier2012} and only recently, the field range has been extended up to 14~T by an NMR study~\cite{Klanjsek2012}. These measurements revealed a minimum of the phase boundary around 8.6~T and suggested that this minimum is related to a rearrangement concerning the magnetic inter-chain ordering. Theoretical studies on the basis of bosonization combined with mean-field theory for the inter-chain interaction suggested an alternative possible origin for a non-monotonic phase boundary, namely a possible change from the incommensurate phase with spins parallel to $c$ towards a transverse staggered phase with spins oriented (mainly) perpendicular to $c$~\cite{Okunishi2007,Okunishi2010}. Despite the drastic broadening of the high-field anomalies, our data clearly support the occurrence of a minimum of the phase boundary around $\approx 8$~T, but our macroscopic data can neither confirm nor refute any of the proposed microscopic changes of the magnetic structure. In our low-temperature magnetostriction data (Fig.~\ref{fig:tadc}c), we do not observe any indication for the occurrence of a field-induced phase transition around 8~T. However, in view of the very small and broad anomalies of $\alpha(T)$ at the boundary to the incommensurate phase, one might also expect only extremely shallow anomalies at a possible rearrangement of the magnetic structure.

\subsection{Magnetic field perpendicular to the \textit{c} axis}

Up to now, the field direction perpendicular to the easy $c$ axis of \baco seemed to be of comparatively little interest. As described above, low-temperature $M(H)$ curves~\cite{Kimura2006} yield a saturation field of about 40~T and the reported phase boundaries in the region of 9~T are contradictory~\cite{He2006A,Zhao2012}. In both studies, the explicit  field direction within the $ab$ plane has not been specified, but our measurements reveal a significant anisotropy between $H\| a_t$ and $H\| [110]$. Such an in-plane anisotropy is in accordance with the tetragonal symmetry of the pa\-ra\-mag\-ne\-tic high-temperature phase and as shown above, the symmetry in the N\'{e}el-ordered phase is even reduced to orthorhombic; although for macroscopic probes this is (at least partly) masked by the twinning of the crystals.

\begin{figure}[t]
	\centering
		\includegraphics[width=0.99\linewidth]{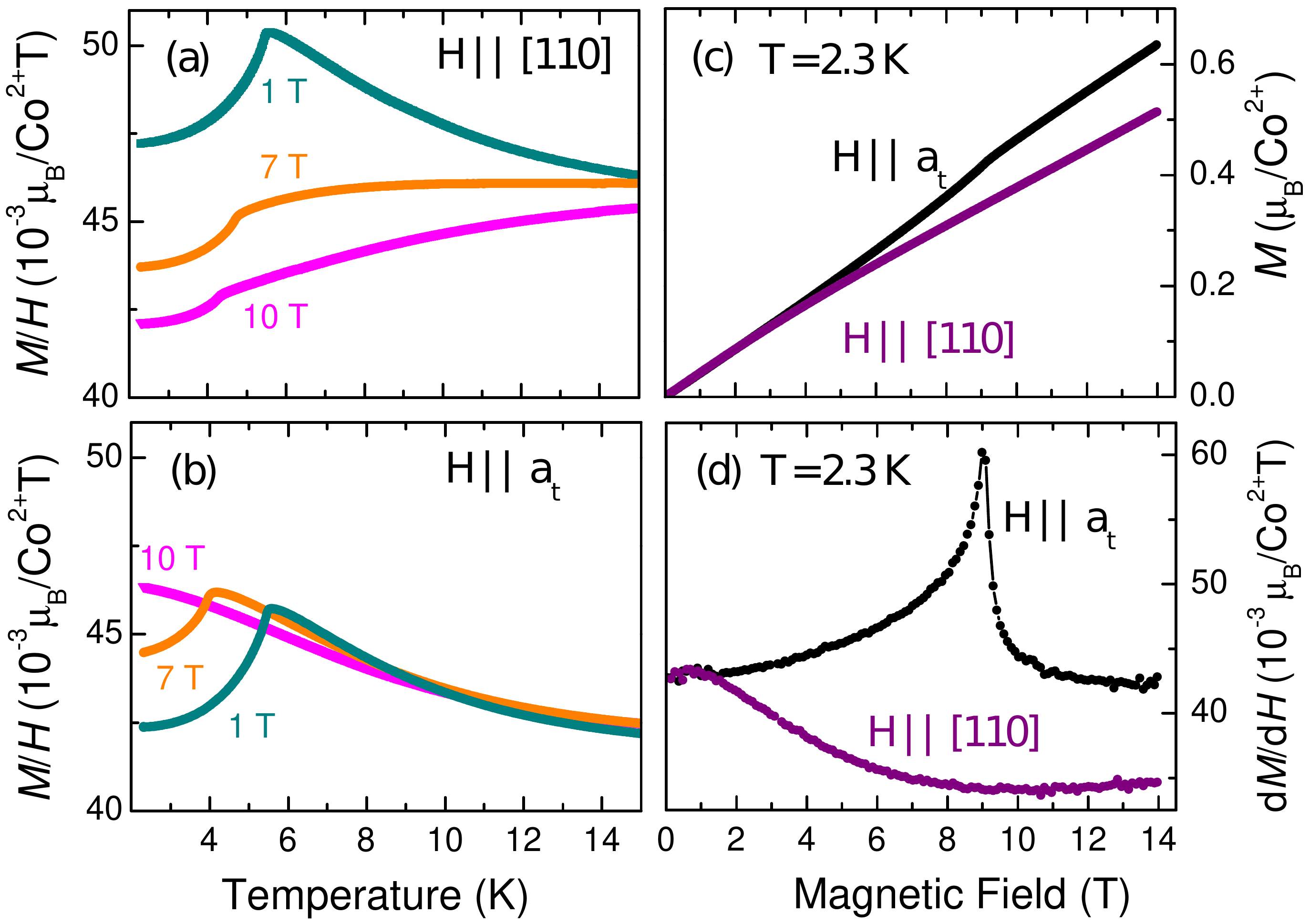}
	\caption{(color online) Left: $T$-dependent magnetization of \baco for different magnetic fields applied either along  [110] (top) or along the tetragonal $a_t$ axis (bottom). Right: Magnetization curves $M(H)$ (top) and the corresponding $\partial M/\partial H$ (bottom) for both field directions, measured at 2.3~K.}
	\label{fig:chi}
\end{figure}

Fig.~\ref{fig:chi} compares representative magnetization data of \baco for $H\| a_t$ and $H\| [110]$. For $H=1$~T, the $M(T)$ curves for both field directions only differ by about 10~\% and show a sharp kink around $T_N \simeq 5.5$~K. With increasing field, however, a pronounced anisotropy develops: Increasing the field to 10~T for $H\| [110]$ causes a continuous suppression of $M/H$ and a weak decrease of \tn, whereas for $H\| a_t$, there is little change of $M/H$ above \tn, but \tn itself is so strongly suppressed that it is no longer visible in the 10~T curve. The right panels of  Fig.~\ref{fig:chi} compare the $M(H)$ curves measured at $T=2.3$~K for both field directions. While $M(H)$ for $H\| [110]$ just continuously increases (with a slight downward curvature) up to the maximum field of 14~T, the $M(H)$ curve for $H\| a_t$ has a clear anomaly at $\simeq 9.2$~T signalling a phase transition as is best seen in the corresponding differential susceptibility $\partial M/\partial H$. A very recent high-field magnetization study has revealed that this phase transition is not related to a saturation of $M$ and has proposed that its microscopic origin lies in the structural arrangement of the CoO$_6$ screw chains~\cite{Kimura2013} (see below). 

In the following, the phase diagrams obtained for the two in-plane field directions are discussed separately. Again, the corresponding phase boundaries are derived by measurements of the thermal expansion and magnetostriction, which do not only have a very high resolution but also yield additional information about the magnetoelastic coupling and a field-induced reorientation of the domains.

\subsubsection{Magnetic field parallel to the [110] direction}

\begin{figure}
		\includegraphics[width=1.0\linewidth]{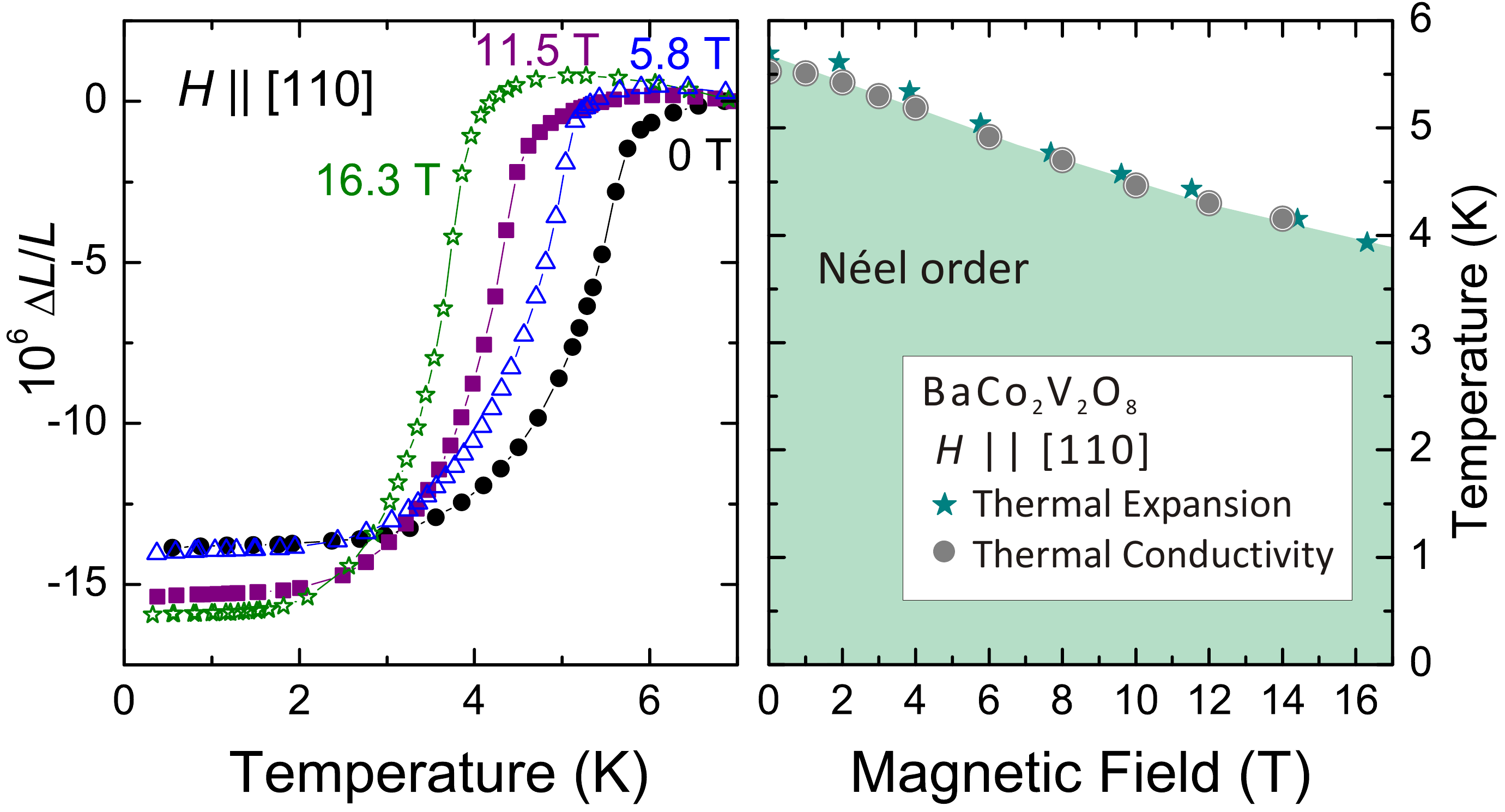} 
	\caption{(color online) Thermal expansion $\Delta L(T)/L$ measured along [110] for various magnetic fields $H \| [110]$ (left) and the corresponding $H-T$ phase diagram (right).}
	\label{fig:tad110}
\end{figure}

Representative measurements of $\Delta L_{\text{[110]}}(T)/L_{\text{[110]}}$ for different fields $H\| [110]$ are displayed in Fig.~\ref{fig:tad110}. In zero field, a spontaneous contraction develops below \tn , which is weakly enhanced on increasing the magnetic field, while \tn , derived from the maxima of the corresponding $\alpha_{\text{[110]}}(T)$ (not shown), continuously decreases with a slope $\partial T_{\text{N}}/\partial H \simeq -0.1$~K/T. Thus, our phase boundary for $H\| [110]$ (Fig.~\ref{fig:tad110}) very well agrees to the $H^{\perp c}-T$ phase diagram based on specific-heat data~\cite{He2006A} and we suppose that those data were measured with $H^{\perp c}\| [110]$. Such an orientation appears natural because the crystal of Ref.~\onlinecite{He2006A} was obtained by spontaneous nucleation and our crystal grown by this method has flat (110) surfaces; see Fig.~\ref{fig:Foto}c.

As discussed, the N\'{e}el ordering causes an orthorhombic splitting  $a_o > a_t > b_o$ and the crystal is twinned below \tn. Such a twinning does, however, not influence the expansion along [110] because both types of domains are rotated with respect to each other by 90$^\circ$ around the $c$ axis and the relative length change along the diagonals of each domain is given by $\Delta L_{110}/L_{110}=\left( \Delta a_o/a_o + \Delta b_o/b_o\right)/2$. The individual $\Delta a_o/a_o$ and $\Delta b_o/b_o$ are of opposite signs (Fig.~\ref{fig:nullfeld}), but we expect their absolute values to be of similar magnitude and not too much larger than the measured $\Delta L_{110}/L_{110} \simeq 1.5\cdot 10^{-5}$. The corresponding orthorhombic splitting $(a_o-b_o)/(a_o+b_o)$ should thus be in the $10^{-4}$ range making it difficult to be detected by standard diffraction techniques as already stated above. 

\subsubsection{Magnetic field parallel to the \textit{a} axis}
\label{sec:Hparallela}

\begin{figure}
	\centering
		\includegraphics[width=1.00\linewidth]{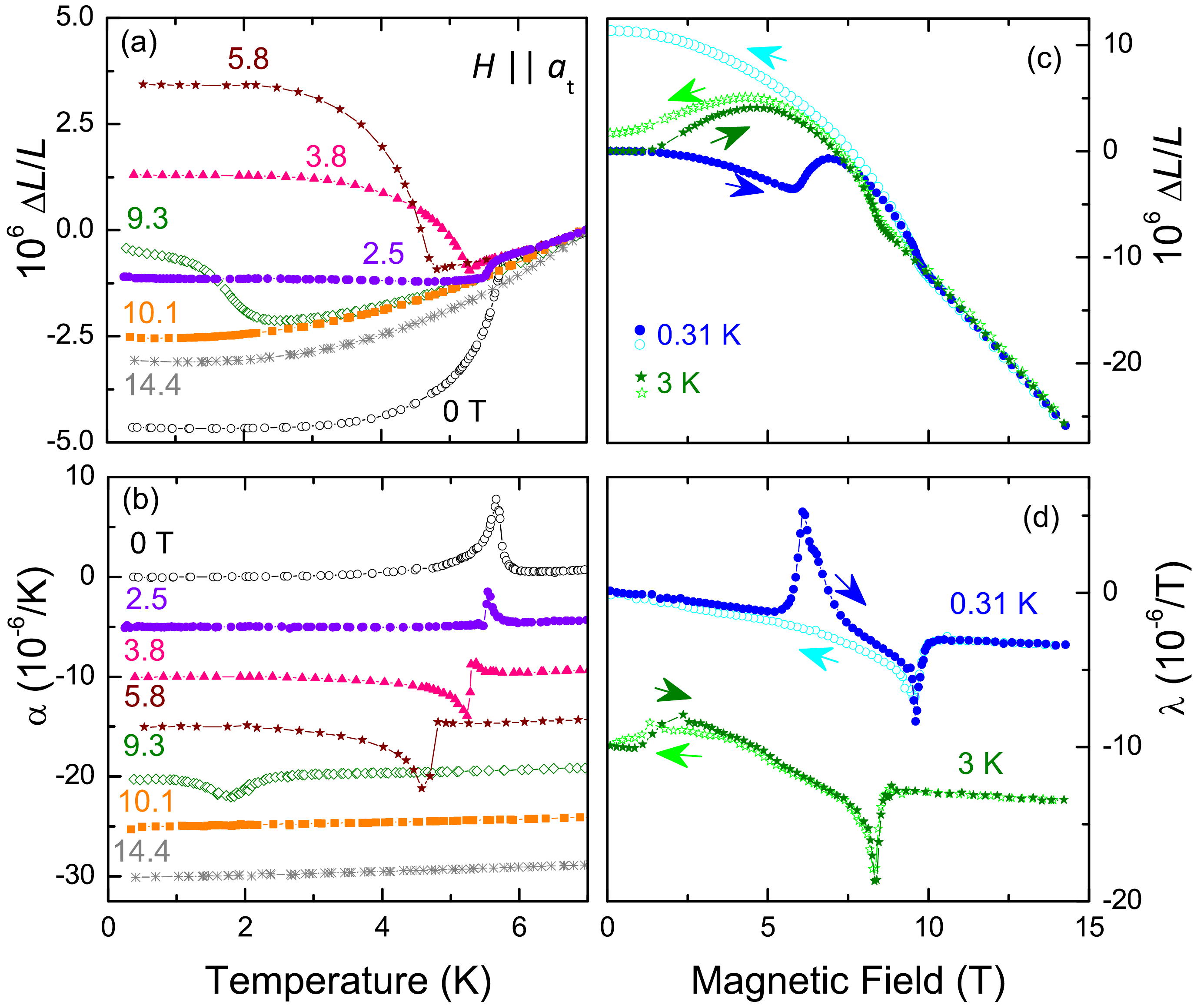}
	\caption{(color online) Thermal expansion (left) and magnetostriction (right) of the tetragonal $a_t$ axis of \baco for magnetic fields $H\|a_t$. The upper and lower panels display $\Delta L/L$ and the corresponding derivatives with respect to $T$ and $H$, respectively. For clarity, the different curves in panels~(b) and~(d) are shifted with respect to each other.   Around 3~T there is a sign change of $\Delta L(T)$ due to a field-induced partial detwinning and  for the same reason the field-dependent data depend on the field-sweep direction (marked by arrows) in the low-field range.}
	\label{fig:tada}
\end{figure}

Now we turn to the configuration with $H$ applied along one of the tetragonal $a_t$ axes. The thermal-expansion data measured along this direction are presented in Fig.~\ref{fig:tada}. In this setup, the domain structure below \tn is such that a small spontaneous contraction is present in zero field. On increasing the magnetic field, the magnitude of the spontaneous contraction rapidly decreases and around 3~T, it even changes sign and turns into a spontaneous expansion, which is of maximum size around 6~T and finally completely vanishes above 10~T. The field dependence of \tn\ is best monitored by considering the shift of the maxima/minima of the corresponding thermal-expansion coefficient $\alpha(T)$ shown in Fig.~\ref{fig:tadc}b. At first glance, the sign change around 3~T might be interpreted as an indication for different magnetic structures above and below that field. However, we already know that the crystal is twinned and a more natural explanation arises from a magnetic-field induced (partial) detwinning by assuming that the magnetic field favours those domains which are oriented with the longer $a_o$ axis parallel to $H$. Thus, if applied along the same $a_t$ axis, a magnetic field and uniaxial pressure would oppose each other because $H$ favors the ``long'' domains being aligned along $H\|a_t$ whereas uniaxial pressure prefers the ``short'' domains along its direction. However, in a transverse configuration, \textit{i.e.} when applied along two orthogonal $a_t$ axes, magnetic field and pressure would support each other because both favor the same type of domain orientation; see Fig.~\ref{fig:nullfeld}. As explained above, the $\Delta L(T,p_i)/L$ data for uniaxial pressure (Fig.~\ref{fig:nullfeld}b) were obtained in a ``pressure-cooled" mode and, analogously, the $\Delta L(T,H)/L$ curves of Fig.~\ref{fig:tada}a were measured in the``field-cooled" mode, \textit{i.e.}\ $H$ has been varied only above \tn. 

The right panels of Fig.~\ref{fig:tada} display some $\Delta L(H)/L$ curves measured at constant $T$ after the sample has been cooled in zero field. At $T=0.31$~K, $\Delta L(H)/L$ shows two anomalies on increasing the field: a step-like increase at $\simeq 6$~T and a slope change at $\simeq 10$~T. The slope change is reversible, whereas the step-like change is absent in the field-decreasing run. At a higher $T=3$~K, broadened but still hysteretic anomalies occur in $\Delta L(H)/L$ around 2~T for both field-sweep directions, while at $\simeq 8.5$~T, the analogous reversible slope change is observed as in the low-temperature curves. The slope changes at higher fields correspond to a field-driven crossing of the $2^{nd}$-order phase boundary, which is already known from the temperature-dependent measurements of the left panels of Fig.~\ref{fig:tada}. The anomalies at lower fields show a similar hysteresis behavior as it is known from hysteresis loops of ferromagnets and the more or less step-like increase on increasing field is, thus, naturally explained by the field-induced domain orientation as discussed above.

Note, however, that in contrast to a ferromagnet, the field-induced orientation of the domains in \baco cannot be  reversed by reversing the magnetic-field direction. Instead one would have to apply either a magnetic field along the perpendicular $a_t$ direction or, alternatively, one could apply uniaxial pressure along the direction of the original field. The possibility of manipulating a certain type of ``ferro-'' domains, \textit{e.g.} ferromagnetic, ferroelectric and/or ferroelastic, by different types of external fields, \textit{i.e.} magnetic or electric fields or uniaxial pressure/strain, is of strong potential interest in the context of data storage, and during the last decade it has attracted a lot of attention in the field of so-called multiferroic materials~\cite{Fiebig,Khomskii}. Typical multiferroic ordering phenomena arise from coupled (anti-)ferromagnetic and ferroelectric order parameters and, in most cases, a more or less complex magnetic structure yields the primary order parameter, while the ferroelectric polarization is a secondary (or improper) order parameter. The coupling of both order parameters, then, allows the manipulation of the magnetization  by an electric field or, \textit{vice versa}, changing the polarization by a magnetic field, and these changes are most pronounced if there are switchable domains. In the present case of \baco, the primary order parameter stems from the (antiferro-)magnetic order while the spontaneous strain is a secondary order parameter and, due to the finite order-parameter coupling, the corresponding domains can be reoriented either by a magnetic field or by uniaxial pressure.

\begin{figure}
	\centering
			\includegraphics[width=0.99\linewidth]{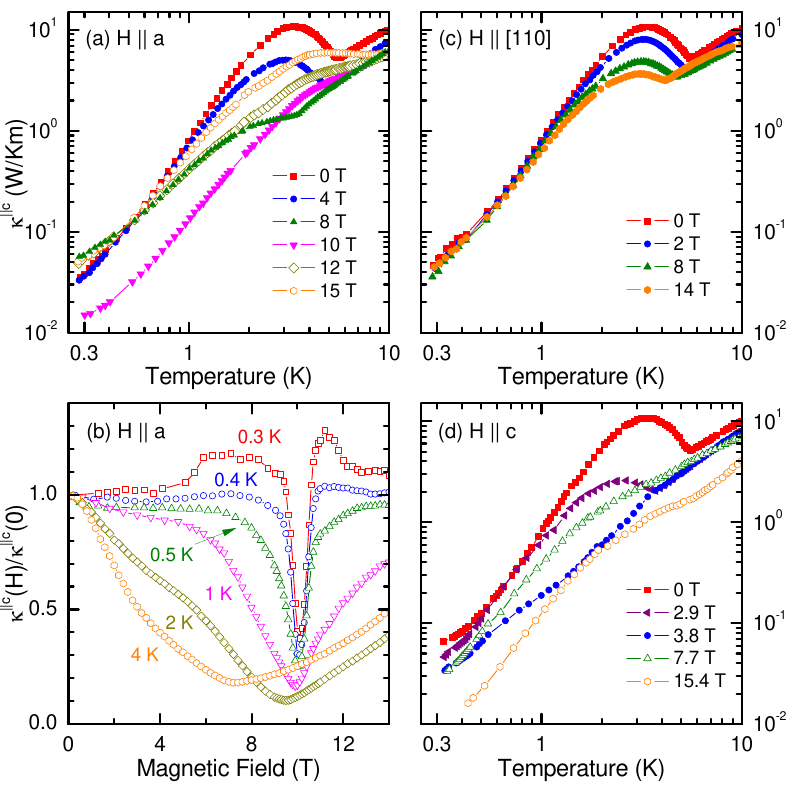}
	\caption{(color online) Panels (a), (c), and (d) display the thermal  conductivity $\kappa^{\|c}(T)$ along the spin-chain direction $c$ of \baco for various  magnetic fields applied along different directions. Panel (b) displays the relative change $\kappa^{\|c}_c(H)/\kappa^{\|c}_c(0)$ for  $H \| a_t$ at various temperatures.}
	\label{fig:kappa}
\end{figure}

Fig.~\ref{fig:kappa} displays representative thermal-conductivitiy data of \baco for different magnetic field directions. In all these measurements, the heat current has been applied along $c$, \textit{i.e.} parallel to the spin-chain direction. One  aim of these measurements was to study whether there is a sizable magnetic contribution to the total heat transport as it is found in several Heisenberg spin-1/2-chain or ladder materials~\cite{SologubenkoRev,hess07}. A standard way to separate a quasi-1D magnetic contribution from the phonon heat transport is to consider the difference of the thermal conductivity along ($\kappa^\|$) and perpendicular ($\kappa^\perp$) to the chain direction. However, our low-temperature data of $\kappa^\perp$ (not shown) obtained with a heat current along [110] are almost identical to the $\kappa^\|(T)$ of Fig.~\ref{fig:kappa}. The lack of anisotropy between $\kappa^\|$ and $\kappa^\perp$ is a strong indication against the presence of a sizable magnetic heat transport in \baco . This finding has recently been published already in Ref.~\onlinecite{Zhao2012} by Zhao \textit{et al.}\ who have drawn exactly the same conclusion. As is seen in Fig.~\ref{fig:kappa}, the zero-field $\kappa(T)$ has a sharp minimum at $T_N \simeq 5.5$~K, which signals strong phonon scattering by magnetic fluctuations around the ordering transition and following the minimum of $\kappa$ for different magnetic fields (and field directions) reveals the respective phase boundaries. The $\kappa(T)$ data for different field directions fully confirm the strongly anisotropic phase diagrams of \baco for $H\| a$, $H\| [110]$, and $H\| c$ as discussed above. Moreover, our $\kappa(T,H)$ data for $H\| a$ and $H\| c$ (Figs.~\ref{fig:kappa}a, b, and d) very well agree to the respective measurements of $\kappa(H,T)$ for  $H\perp c$ and $H \| c$ of Ref.~\onlinecite{Zhao2012}. Because the additional in-plane anisotropy was unknown, Zhao \textit{et al.} interpreted the deviation of their $H^{\perp c}$-phase boundary from the previously known $H^{\perp c}$-phase boundary~\cite{He2006A} as indication for the occurrence of a new phase around 10~T. This discrepancy is, however, due to a mixing up of the different phase diagrams for $H\| [110]$ and $H\| a$  and is fully resolved by considering the very different results of $\kappa(T,H)$ for the different in-plane field directions $H\perp c$, see Figs.~\ref{fig:kappa}a and \ref{fig:kappa}c. 

\begin{figure}
	\centering
		\includegraphics[width=0.99\linewidth]{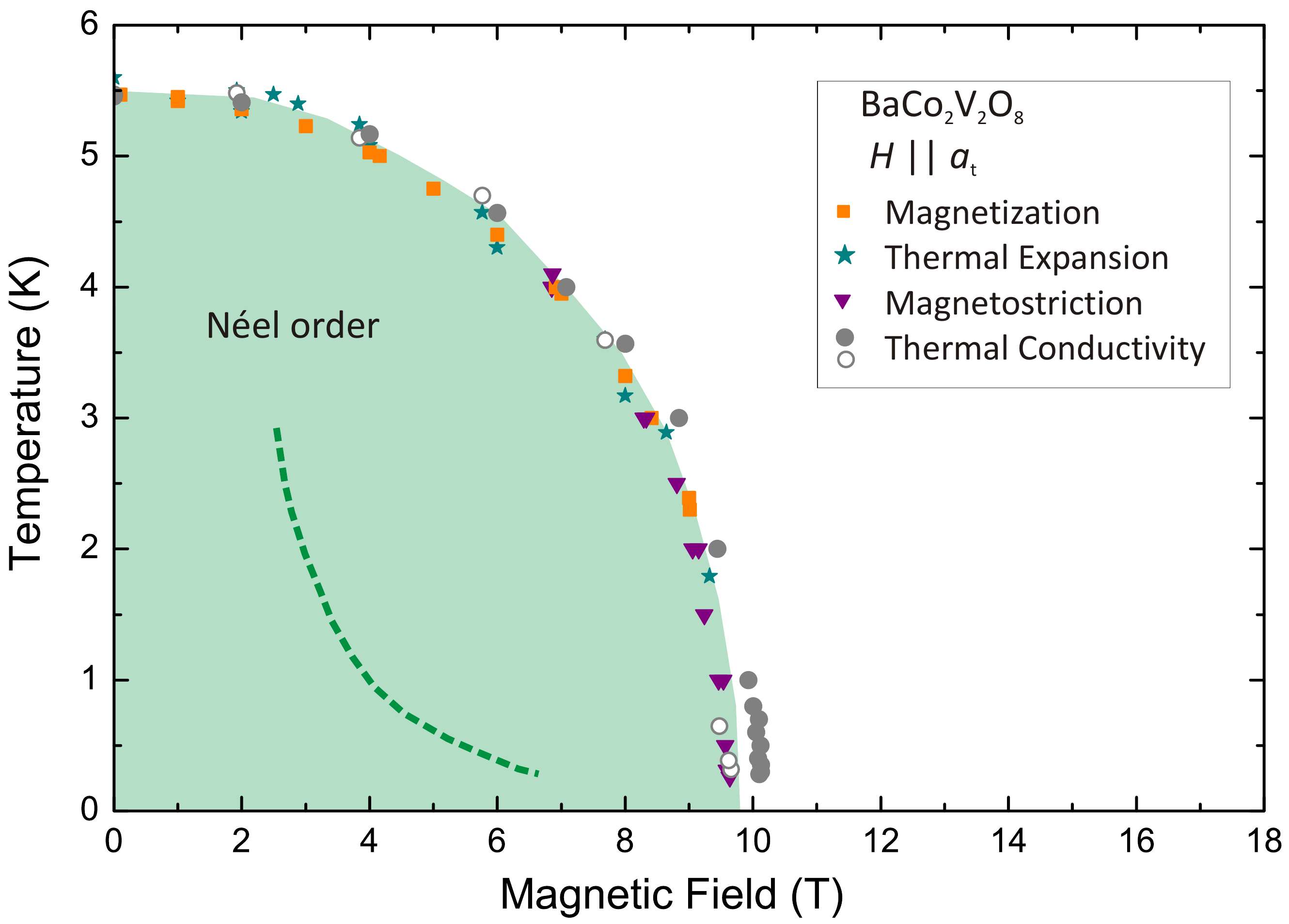}
	\caption{(color online) Phase diagram of \baco for $H\| a_t$. The dashed line marks the region of the field-induced reorientation of magnetic domains.}
	\label{fig:phadia}
\end{figure}

The phase diagram of \baco for magnetic fields $H\| a$ is given in Fig.~\ref{fig:phadia}. The shape of this phase boundary is consistently confirmed by all the different probes. The small quantitative deviation of the critical fields obtained by one set of thermal-conductivity measurements compared to the other data most probably arises from a slight misalignment of the magnetic-field direction. We mention that a similar strong suppression of \tn\ for $H\perp c$ has been reported for the closely related material \srco in Ref.~\onlinecite{PhysRevB.73.212406}. Because \srco and \baco have essentially the same structure and also almost the same N\'{e}el temperature in zero field, their magnetic subsystems should be very similar and we expect that a study of the in-plane anisotropy of \srco would reveal very similar results as obtained here for \baco. As indicated by the dashed line in Fig.~\ref{fig:phadia}, the partial field-induced detwinning of the sample causes some hysteresis behavior, which not only depends on the actual path in the $H-T$ plane but is also sample dependent because it is probably influenced by internal strains and/or defects. Concerning the actual phase boundary, however, we did not observe hysteresis effects in any of our measurements identifying the corresponding transition as a continuous 2$^{nd}$-order phase transition.  Thus, our data strongly  suggest that \baco has a quantum critical point at $H_{crit}^{\|a}\simeq 10$~T. 

As can be seen in Fig.~\ref{fig:tada}b, the thermal expansion coefficient $\alpha$ also changes sign at 10~T, which is a typical signature of a quantum phase transition~\cite{Zhu2003,Garst2005}. For purely one-dimensional magnets, this sign change can even take place in form of a pole, {\it i.e.} $\alpha(T\rightarrow 0)$ diverges~\cite{lorenz2008,Rohrkamp2010}, whereas for higher-dimensional systems, such a sign change via a divergence is only present for the Gr\"uneisen ratio $\alpha/C$ ($C$ is the specific heat) and the magnetocaloric effect $\partial \ln T/\partial H$~\cite{Garst2005,Kuechler03,Kuechler04,Lorenz2007,Tokiwa2009}. In the present case of \baco , we are most probably dealing with a three-dimensional quantum phase transition from the N\'{e}el-ordered phase to a spin-liquid state. Very recently, this transition has also been observed by high-field magnetization data and the strong in-plane magnetic-field anisotropy of \baco has been explained by a structural peculiarity of the Co$^{2+}$ screw chains arising from a slight distortion of the CoO$_6$ octahedra~\cite{Kimura2013}. According to this model, a homogenous magnetic field $H\| a$ is accompanied by a staggered transverse field component via a staggered $g$ tensor, whereas no such transverse component occurs for $H\| [110]$. As a consequence, this model predicts a collapse of the N\'{e}el order at already rather small fields of $\simeq 7$~T applied along the $a$ direction.

\begin{figure}[t]
	\centering
		\includegraphics[width=0.99\linewidth]{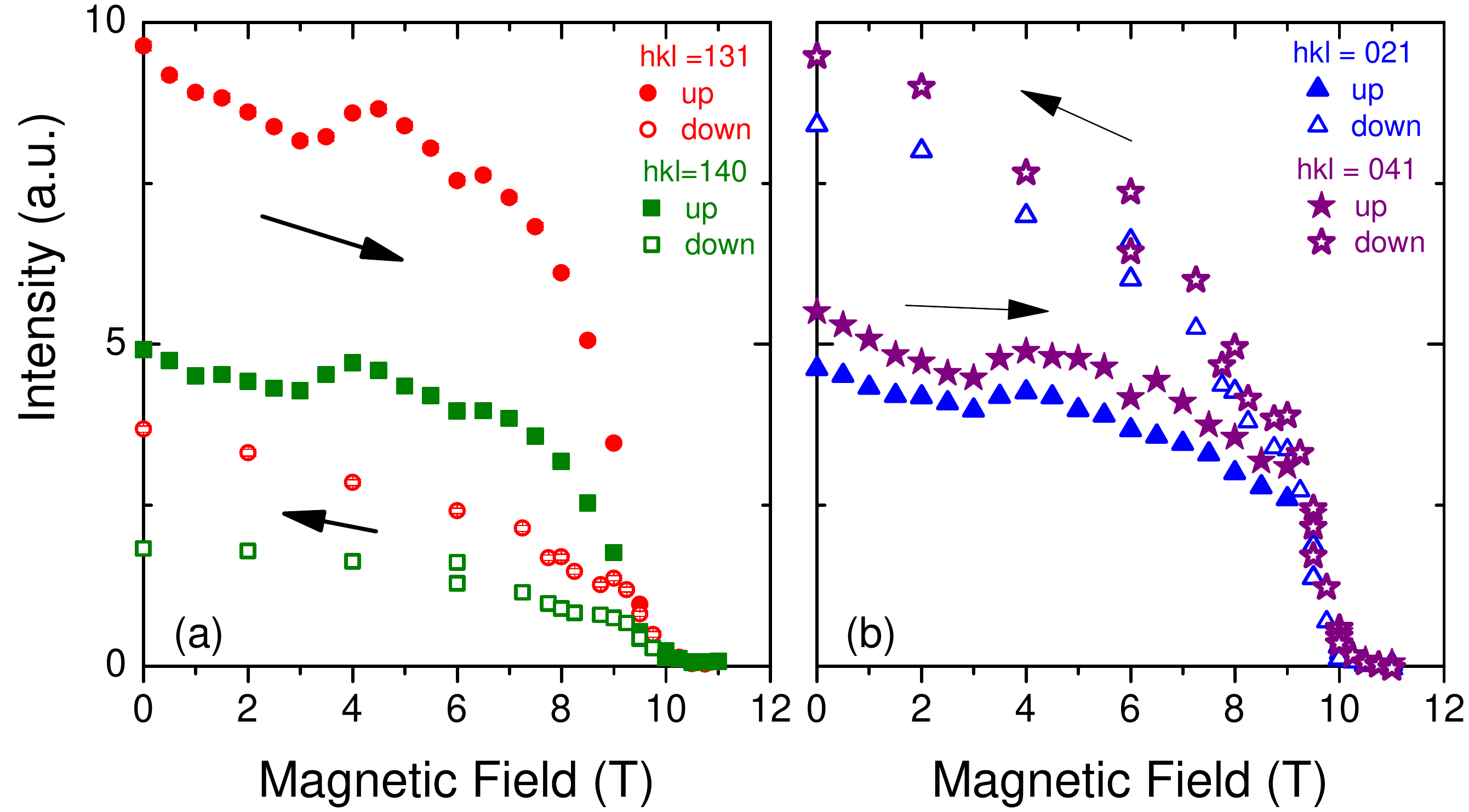}
	\caption{(color online) Intensities of several magnetic Bragg peaks of the N\'{e}el-ordered phase of \baco as a function of magnetic fields $H\| a_t$ at $T\simeq 50$~mK. The peaks of panel (a) arise from one type of domains while those of (b) represent the other type~\cite{Grenier2012}. The different intensities obtained with increasing or decreasing magnetic field, shown as solid and open symbols, respectively, proof a field-induced reorientation of the magnetic domains in the ordered phase.}
	\label{fig:neutron}
\end{figure}

As is shown in Fig.~\ref{fig:neutron}, this hypothesis is verified by our low-temperature ($T\simeq 50$~mK) single-crystal neutron diffraction data. The field dependent intensities of several magnetic superstructure reflections of the zero-field N\'{e}el-ordered phase collapse at $H_{crit}^{\|a}\simeq 10$~T. However, the suppression of the magnetic intensity is not homogeneous and there is a significant hysteresis. Moreover, all reflections show a clear anomaly around 4~T, whose origin remains unclear. Based on the present data, it cannot be excluded that this anomaly is related with the domain structure and extinction effects. As discussed above, the zero-field N\'{e}el phase contains two types of magnetic domains, which are represented by the different magnetic superstructure peaks shown in Fig.~\ref{fig:neutron}a and ~\ref{fig:neutron}b, respectively (see Ref.~\onlinecite{Grenier2012}). On increasing the magnetic field, the the peak intensities of panel~(a) rapidly decrease already above $\simeq 9$~T, while those of panel~(b) only drop close to 10~T. Such a different field dependence is naturally explained by a field-induced domain reorientation close to the critical field $H_{crit}^{\|a}\simeq 10$~T. Moreover, the fact that one type of domains is favored by the magnetic field $H\| a_t$ also explains why the field-decreasing run results in a different intensity distribution as compared to the original zero-field data. The suppression of the intensities of all measured peaks above $H_{crit}^{\|a}\simeq 10$~T implies that either the high-field phase is indeed a spin-liquid state or that another type of magnetic order settles above $H_{crit}^{\|a}$. Because the present data set does not yet allow an unambiguous conclusion, an additional neutron scattering study of the high-field phase is needed to look whether some magnetic signal appears elsewhere in the reciprocal lattice or not.

\section{Summary}
\label{sec:Sum}

We have presented a detailed study of the magnetic phase diagrams of \baco for field directions parallel and perpendicular to the easy axis $c$. For a magnetic field along the easy axis, our data essentially confirm previous results, in particular the non-monotonic field dependence of the transition temperature to the incommensurate phase~\cite{Klanjsek2012}.  The phase diagram for magnetic fields applied within the $ab$ plane is more complex than originally expected. The zero-field  N\'{e}el ordering is only weakly suppressed by a magnetic field along [110]. However, applying the magnetic field along the tetragonal $a_t$ axis, the phase diagram shows a strong field dependence and the  N\'{e}el temperature vanishes above a critical magnetic field $H_{crit}^{\|a}\simeq 10$~T. Our thermal-expansion data indicate that this quantum critical point signals a quantum phase transition from the three-dimensionally ordered N\'{e}el phase to a spin-liquid state. Based on magnetization data and an analysis of the microscopic details of the structure of \baco , a collapse of the N\'{e}el order for this field direction has been predicted recently~\cite{Kimura2013} and is experimentally supported by our neutron scattering data. 

Furthermore, our high-resolution measurements of the thermal expansion show that the N\'{e}el ordering in \baco is accompanied by a structural transition from the tetragonal to an orthorhombic structure. This reduction of the structural symmetry can be straightforwardly explained by a finite magnetoelastic coupling of the inter-chain couplings and the fact that the magnetic symmetry is lower than tetragonal. The corresponding lattice distortions are rather weak and the macroscopic crystals are heavily twinned with two types of $90^\circ$~domains. The domain orientation in the ordered phase can be influenced externally, either by applying uniaxial pressure parallel $a_t$ or by a magnetic field along the $a_t$ axis.

\begin{acknowledgments}
We acknowledge helpful discussions with M.~Garst and thank E.~Ressouche for assistance during the neutron scattering experiments. This work was supported by the Deutsche Forschungsgemeinschaft through SFB 608.
\end{acknowledgments}


\begin{thebibliography}{10}
\bibitem{remark}%
  \BibitemOpen
  \bibinfo {note} {As a consequence of the alternating handedness of the
  neighboring screw chains, different inter-chain couplings $J^\perp_{NN}$ and
  $J^\perp_{NNN}$ are present in BaCo$_2$V$_2$O$_8$. Moreover, $J^\perp_{NN}$
  and $J^\perp_{NNN}$ are not oriented exactly parallel to the $a_t$ axes and the 
  $\langle 110 \rangle$ directions, but these directions are
  symmetry axes of $J^\perp_{NN}$ and $J^\perp_{NNN}$, respectively.}%
  \bibAnnoteFile{Stop}{remark}%
\end{thebibliography}
\end{document}